\documentclass[10pt,preprint]{aastex}

\newcommand\asec{^{\prime\prime}}
\newcommand\etal{{\it et al.}}
\newcommand\lae{\mathrel{<\kern-1.0em\lower0.9ex\hbox{$\sim$}}}
\newcommand\gae{\mathrel{>\kern-1.0em\lower0.9ex\hbox{$\sim$}}}

\newcommand\mone{$^{-1}$}
\newcommand\mtwo{$^{-2}$}

\begin{document}
\title{Resolving the Shocks in Radio Galaxy Nebulae: 
HST and Radio Imaging of 3C~171, 3C~277.3, and PKS~2250-41
\footnote{Based on observations made with
the NASA/ESA Hubble Space Telescope, obtained at the Space Telescope Science
Institute, which is operated by the Association of Universities for Research
in Astronomy, Inc., under NASA contract NAS 5-26555. These observations are
associated with program 6657 (PI C. Tadhunter).}}

\author{Avanti Tilak\altaffilmark{a}, Christopher P. O'Dea\altaffilmark{b},Clive Tadhunter\altaffilmark{c}, Karen Willis\altaffilmark{c}, Rafaella Morganti\altaffilmark{d}, Stefi A. Baum\altaffilmark{d}, Anton M. Koekemoer\altaffilmark{f}, and Daniele Dallacasa\altaffilmark{g,h}}
\altaffiltext{a}{Department of Physics \& Astronomy, The Johns Hopkins University, 3400 N. Charles Street, Baltimore, MD-21218;{tilak@pha.jhu.edu}}
\altaffiltext{b}{Department of Physics, Rochester Institute of Technology,One Lomb Memorial Drive, Rochester, NY-14623;{odea@cis.rit.edu}}
\altaffiltext{c}{Department of Physics \& Astronomy, University of Sheffield, Western Bank, Sheffield, S10 2TN, UK;{C.Tadhunter@sheffield.ac.uk}, {K.Willis@sheffield.ac.uk}}
\altaffiltext{d}{Netherlands Foundation for Research in Astronomy, Postbus 2, 7990 AA Dwingeloo, The Netherlands;{morganti@astron.nl}}
\altaffiltext{e}{Center for Imaging Sciences, Rochester Institute of Technology, One Lomb Memorial Drive, Rochester, NY-14623; {baum@cis.rit.edu}} 
\altaffiltext{f}{Space Telescope Science Institute\footnote{Operated by the 
Association of Universities for Research in Astronomy under contract
NAS 5-26555 with the National Aeronautics and Space Administration.}, 3700 San Martin Dr., Baltimore, MD 21218; {koekemoe@stsci.edu}}
\altaffiltext{g}{Dipartimento di Astronomia, via Ranzani 1, I-40127 Bologna, Italy}
\altaffiltext{h}{Ist. di Radioastronomia, CNR, Via Gobetti 101, I-40129 Bologna, Italy;{ddallaca@ira.cnr.it}}
%\author{draft:\today}

%\clearpage

\begin{abstract}
\noindent
We present the results of HST/WFPC2 medium and narrow band imaging and VLA and
MERLIN2 radio imaging of three powerful radio galaxies: 3C 171, 3C 277.3, and
PKS 2250-41. We obtained images of the rest frame [OIII]$\lambda$5007 and
[OII]$\lambda$3727 line emission using the Linear Ramp Filters on WFPC2.
The correlations between the emission line morphology and
the [OIII]/[OII] line ratios with the radio emission seen in ground based
observations are clarified by the HST imaging. We confirm that the radio lobes
and hot-spots are preferentially associated with lower ionization gas.
3C 171 exhibits high surface brightness emission line gas mainly along
the radio source axis. The lowest ionization gas is seen at the Eastern hot
spot. In 3C 277.3 there is bright high ionization gas (and continuum) offset 
just to the east of  the  radio knot K1. Our observations are consistent 
with previous work suggesting that this emission is produced by precursor 
gas ionized by the shock being driven into the cloud by the 
deflected radio jet. In PKS 2250-41 we resolve the emission line arc which 
wraps around the outer rim of the western lobe. The lower ionization [OII] 
emission is nested just interior to the higher ionization [OIII] emission 
suggesting that we have resolved the cooling region behind the bow shock. 
We also detect possible continuum emission from the secondary hot-spot.
Thus, our observations support the hypothesis that in these sources, the
interaction between the expanding radio source and the ambient gas strongly
influences the morphology, kinematics,  and ionization of the gas.
\end{abstract}


\section{Introduction}

\noindent
The association of powerful radio galaxies with luminous extended 
emission line nebulae is well established (e.g., Heckman \etal\ 1982, 
Meisenheimer \& Hippelein 1992, Baum \& McCarthy 2000).
There has been great debate over whether nuclear
photo-ionization or shock heating is the dominant source of ionization.
However, it is becoming clear that both mechanisms are present in the
extended nebulae, e.g., \cite{cla98,vill99,ode02}. In principle 
it should be possible to use the properties of the shocked gas to study 
the nature of the jet-cloud
interaction and constrain the physics of the propagation of the radio source. 
Here we present a high resolution HST/WFPC2 imaging study of three
powerful radio galaxies in which previous ground based imaging and
spectroscopy has revealed the presence of shocked gas in the extended
nebula (Clark \etal\ 1997, 1998; Villar-Mart\'{\i}n \etal\ 1999; Tadhunter
\etal\ 2000; Solorzano-I\~narrea \& Tadhunter 2003).
We detail the data reduction and analysis techniques in \S2. The properties of the 
sources are summarized in Tables 1 \& 2. In \S 3 we present the results of 
the emission line imaging and a comparison with the radio source morphology. 
The results are summarized in \S 4. 
Throughout the paper, we have used H$_0$=75 km s\mone Mpc\mone and q$_0$=0.0.

\section{Observations and Data Analysis}

\subsection{Emission Line Images}

\noindent
We observed the redshifted [OIII]$\lambda$5007 and [OII]$\lambda$3727 emission
lines using the HST/WFPC2  Linear Ramp Filters 
(LRF). The LRF are divided into four parallel strips each with the central 
wavelength varying across each strip by about 6 percent \citep{bir02}. In 
general for a given wavelength, the effective field of view is about 
13\arcsec. The object was placed in that part of the chip which corresponds 
to the desired wavelength. In addition to the narrow-band 
emission line images taken with the ramp
filters, intermediate-band continuum images were taken using the
F467M and F547M filters, in order to facilitate accurate continuum
subtraction and allow examination of the detailed continuum structures.
The central wavelengths and bandwidths of the medium band filters were chosen 
to avoid
bright emission lines. The parameters of the HST observations are given 
in Table 3. 

\noindent
The data were calibrated with standard software in STSDAS. Due to the fine 
vignetting of the LRF filters, flats obtained from nearby narrow band filters
were used for flat fielding (Biretta, Baggett \& Noll 1996). The DRIZZLE algorithm 
\citep{fru02, koe02} was used for combining dithered images. 
The algorithm is also very efficient at rejecting cosmic rays. These dithered
images were used in the subsequent analysis. The tasks in the SYNPHOT package 
were used for flux calibration. Adopting an input ``average'' emission line 
profile, we used CALCPHOT to estimate the number of counts expected for a 
Gaussian signal. The characteristics of this input were then used to convert 
the image from counts to flux units. The photometric calibration of the LRF
data should be accurate to $\sim 3\%$ (Biretta \etal\ 1996). 

\noindent
It was necessary to register the LRF and continuum images using WREGISTER, 
since the source is not located at the same pixels in these images. 
This task performs a spatial transformation so that the WCS 
coordinates of the input image match the coordinates of the reference image 
at the same logical pixel coordinates. This first shift is good to a few pixels.
We then manually shifted by a few pixels to register the nucleus on the images.  
Our images do not show significant dust, so the location of the nucleus was
obvious on the continuum and line images. 

\noindent
We used CALCPHOT to estimate the flux expected through each of the filters 
for a standard elliptical galaxy template. We used the ratios of the estimated
fluxes  to scale the medium band image to subtract the continuum contribution
from the LRF images. Over much of the extent of the nebulae, the
continuum is weak or undetected and the stellar continuum correction 
to the LRF images ranges from a few percent  to less than one percent 
of the counts in the LRF image. In a few locations, e.g.,
the nuclei of 3C277.3 and PKS2250-41, the correction to the LRF image
is  as large as 10-20\%.   In principle, there could be a contribution 
to the continuum from e.g., scattered nuclear light or a young stellar
population (e.g., Chambers \& Miley 1990). 
If this were the case, the result could be an increase in
the correction to the LRF images, but over most of the nebula
the correction is still a small fraction of the counts in the LRF image. 

\subsection{Radio Images}

\noindent
Radio observations images for all three sources  were obtained with the VLA
in the A configuration. In addition a 36 hour observation on 3C 277.3 was 
obtained using MERLIN2 (Table 4). Calibration and data reduction was 
achieved using standard techniques in the NRAO AIPS software.
Observations of a bright point source calibrator were
used to determine the antenna gains and phases and to place the observations
in the radio reference frame with an accuracy of $\sim 0.1\asec$. The flux
density scale was determined using observations of 3C286. After
calibration, the data were deconvolved using the CLEAN algorithm, and
the instrumental gains and phases were further corrected using
self-calibration. 
The VLA and MERLIN2 data for  3C277.3 were combined to produce a final image.

\subsection{Data Analysis}

\noindent
The average signal for the background for each of the emission line images was
 obtained. This background level was subtracted from the corresponding images. 
The emission line images were smoothed with a Gaussian with a FWHM of 2 pixels 
in order to increase the surface 
brightness sensitivity. The smoothing also mitigates the consequences 
of any small errors in the image registration.
The pixels with values less than 3$\sigma$ were masked. The masked [OIII] 
image was divided by the masked [OII], to obtain the ratio image. 
In pixels where the [OII] flux was masked, but [OIII] was not, we divided the [OIII] 
flux by the 3$\sigma$ value for the unmasked [OII]
image so as to obtain a lower limit on the ratio in those regions.

\noindent
The optical images were rotated so that they were aligned with the radio 
images and then the optical and radio nuclei (which are visible on all images)
were registered and the images overlayed. 
We estimate that the radio and optical images are registered to 
within 1 pixel ($0.05\arcsec$). The radio and optical images and 
various overlays are presented in Figures 1-12.

\noindent
A small but non-negligible redenning correction \cite{vei87,ost89}, was 
applied for all three galaxies (Table 1). For 3C 171 and 
PKS 2250-41, the reddening correction includes Galactic as well as internal 
contributions and was determined using observed 
H$\alpha$/H$\beta$ ratios \citep{cla97, cla98}. 
For 3C 277.3 the H$\alpha$/H$\beta$ and H$\gamma$/H$\beta$  
measured by Solorzano-I\~{n}arrea \& Tadhunter (2003), are consistent
with Case B suggesting that the redenning is negligible. 

\section{RESULTS }

\subsection{3C 171}

\noindent
3C 171 is a well studied FR II radio source at redshift of 0.2384 (e.g., 
Heckman \etal\ 1984, Baum \etal\ 1988, Blundell 1996, Clark \etal\ 1998, 
Tadhunter \etal\ 2000, Solorzano-I\~{n}arrea \& Tadhunter  2003). The radio 
image shows a faint nucleus and two hot-spots and considerable extended 
diffuse emission which includes large `plumes' which extend roughly 
perpendicular to the radio lobes \cite{blun96}.
Heckman \etal\ (1984) and Baum \etal\ (1988)
found that the inner structure of the powerful radio source (within the 
hotspots) 3C~171 is co-spatial with an 
extended emission line nebula. 
Tadhunter \etal\ (2000) obtained deep H$\alpha$ images which revealed that 
there is faint 
diffuse gas extending beyond the radio source and that the morphology of the
emission line gas is closely related to the radio morphology, but does
not exhibit the ``ionization cone" morphology expected for simple AGN
photo-ionization. Long slit \cite{cla98} and integral field spectroscopy 
\cite{solo03} revealed (1) complex kinematics in the gas including 
line splitting, and (2) lower ionization
gas associated with the hot-spots, and (3) an anti-correlation between line
width and ionization state in the gas. Hardcastle (2003) suggested that the 
depolarization
of the radio emission is produced by a warm shocked medium. 
Taken together, these observations are consistent with the emission line gas
having been shocked by the expanding radio source, e.g. Clark \etal\ (1998).

\begin{figure}
\epsscale{0.85}
\plotone{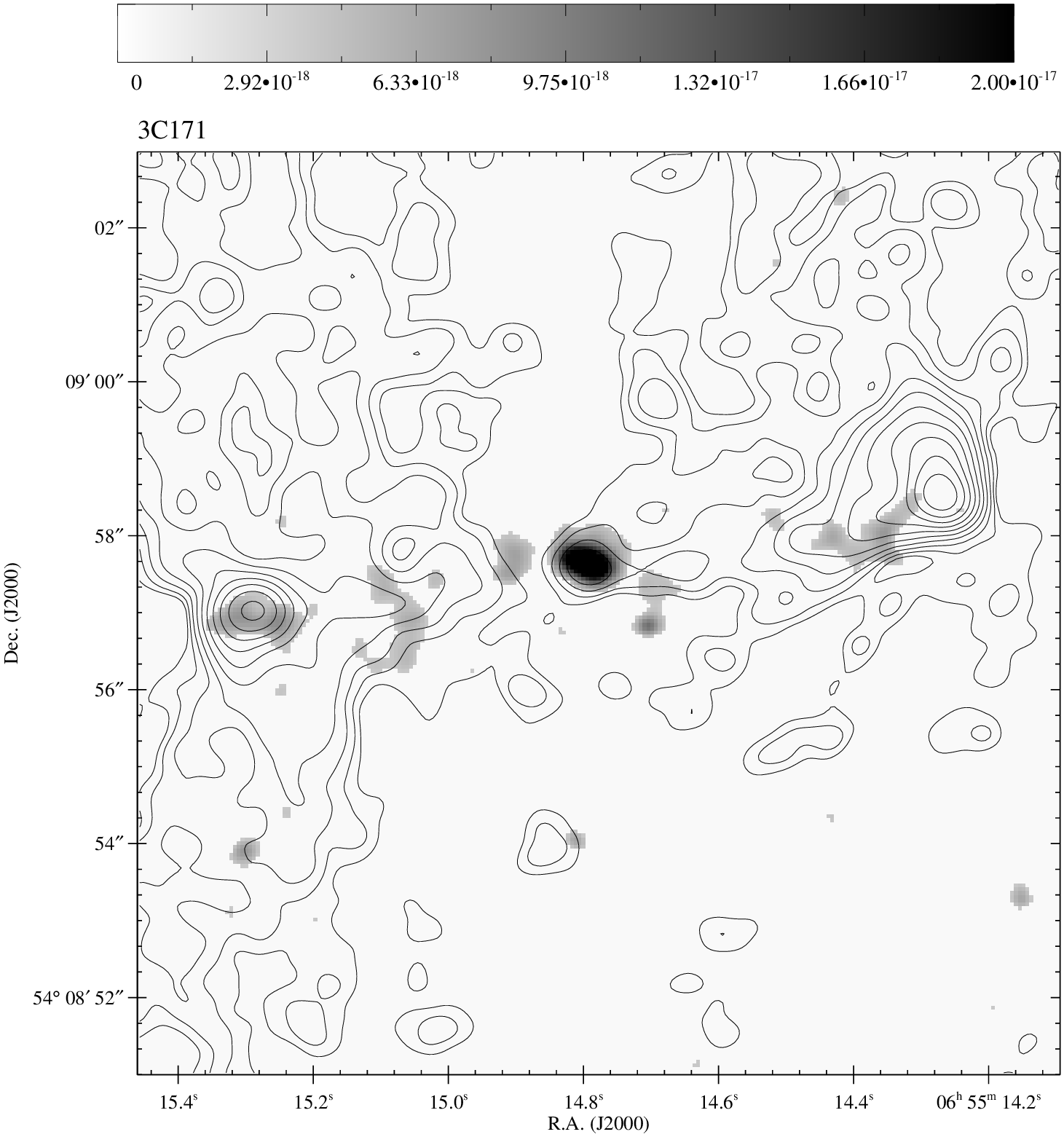}
%\plotone{3C171OII_radio.eps}
\caption{3C171: Contour plot of radio image overlaid on gray-scale surface 
brightness plot of [OII]$\lambda$3727 emission line image. The [OII] image 
has been masked to retain emission above 3.888$\times$$10^{-18}$ ergs s\mone\
 cm\mtwo\ arcsec\mtwo.
The 8 GHz VLA radio image is displayed with contour levels starting at 3 times
the rms noise:
3$\times$2.167$\times$$10^{-5}$$\times$
[1, 2, 4, 8, 16, 32, 64, 128, 256, 512, 1024, 2048]Jy/beam.
\label{171_OII_radio}
}
\end{figure}

\begin{figure}
\epsscale{0.85}
%\plotone{3C171OIII_radio.eps}
\plotone{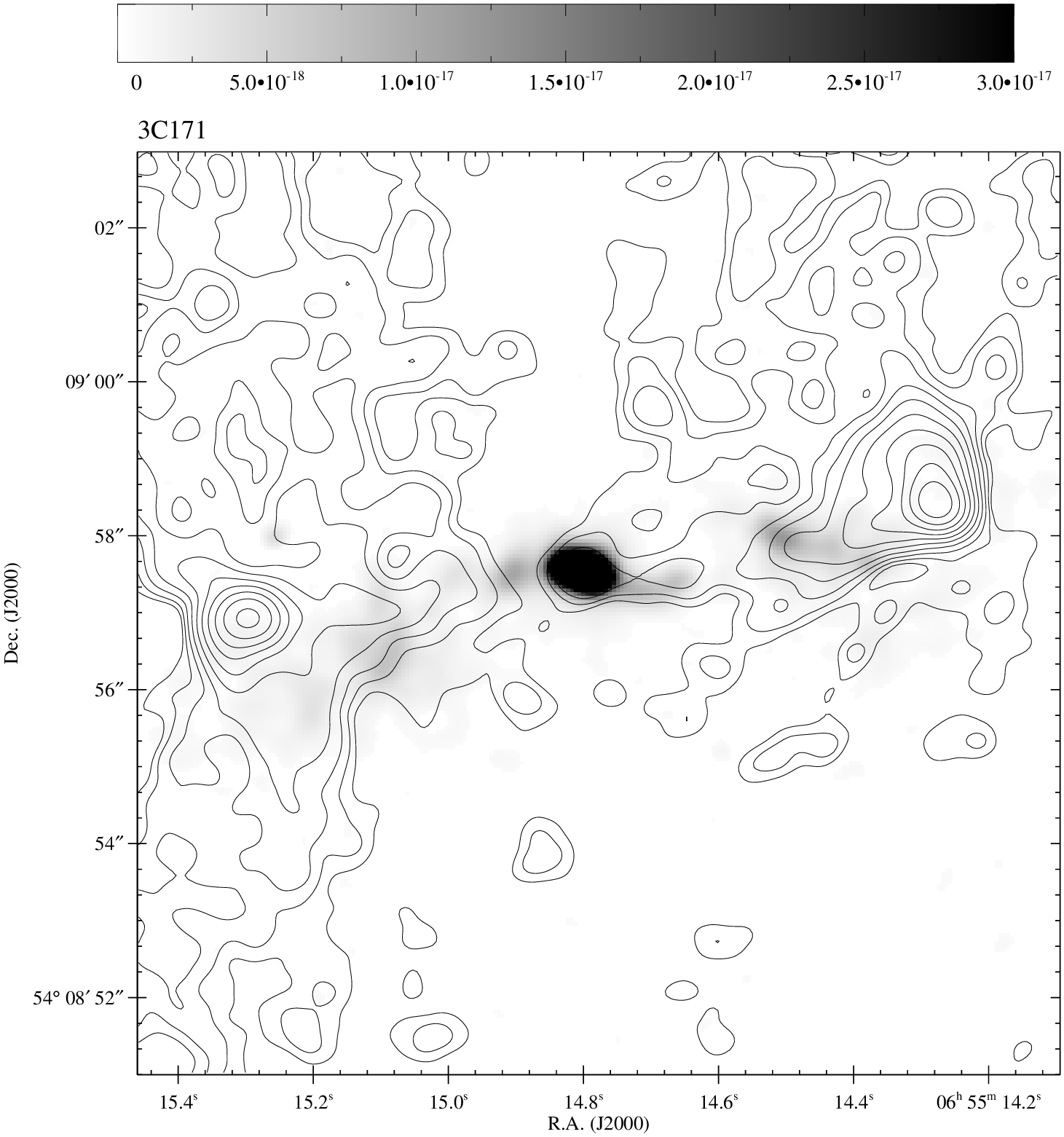}
\caption{3C171: Contour plot of radio image overlaid on gray-scale surface 
brightness plot of [OIII]$\lambda$5007 emission line image. The [OIII] image
has been masked to retain emission above 6.066$\times$$10^{-19}$ ergs 
s\mone\  cm\mtwo\ arcsec\mtwo. 
The 8 GHz VLA radio image is displayed with contour levels starting at
3 times the rms noise: 
3$\times$2.167$\times$$10^{-5}$
$\times$[1, 2, 4, 8, 16, 32, 64, 128, 256, 512, 1024, 2048]Jy/beam.
\label{171_OIII_radio}
}
\end{figure}

\noindent
We present our WFPC2 images and radio images in Figures 1-4. 
Values of the [OIII]/[OII] line ratio are given in Table 6 
for selected regions. The medium band continuum image shows some 
very faint diffuse emission aligned along the radio axis. Comparison 
of the HST and radio images (Figures \ref{171_OII_radio}, 
\ref{171_OIII_radio}) shows that the high surface brightness 
emission line gas tends to follow the central 
ridge line of the radio source. However, there are some 
asymmetries in the source. We detect
[OIII] but not [OII] above a level of $4\times 10^{-18}$ ergs s\mone\
cm\mtwo\ arcsec\mtwo\  associated with the Western hot spot (the [OII] emission 
line gas begins just interior to the hot-spot). However, the lower
resolution ground based observations by Solorzanno-I\~{n}arrea \& Tadhunter 
(2003) show traces of [OII], [OIII] and H$\beta$ line emission in this region. 
On the Eastern side there is [OII] in a band across the hot spot, while the  [OIII] 
extends mainly south of the lobe. We find low ratios of [OIII]/[OII] $\lae$0.5 
associated with the gas near the hot-spots, and higher values $\sim 1.5-2$
 about midway between the lobes and the nucleus. 
The extended structures in this object show a marked similarity with the aligned 
structures in high redshift radio galaxies.

\noindent 
The lack of any line emission for the western hot-spot is surprising,
given that the radio morphology suggests that the jet has hit something
there \cite{blun96}. One explanation might be that the shock 
driven through the warm/cool gas at the western hot-spot may be so 
strong/fast that the gas is heated to high enough temperatures 
that the cooling time becomes long (i.e. the gas has not yet cooled 
to $\sim$1-2$\times$10$^{4}$ K, so that the optical emission lines
become visible). 

\noindent
The anti-correlation between line width and ionization noted by Clark 
\etal\ (1998) suggests that the compressed, high velocity post-shock gas has
a low ionization state, whereas the kinematically quiescent gas has a high
ionization state. It is likely that the high surface brightness regions of
[OII] emission map the kinematically disturbed post-shock gas, whereas
the [OIII] emission is a combination of mostly precursor and some post-shock 
gas. The striking differences between the emission line structures
of the two lines may therefore reflect differences in the spatial
distributions of precursor and cooling post-shock gas; in  this context
it is not surprising that the [OII] is stronger in the eastern lobe than
the [OIII], and that [OIII] dominates in the AGN-photoionized nuclear regions.

\subsection{3C 277.3 (Coma A)}

\noindent
Coma A is classified as intermediate between FRI and FRII radio galaxies, 
displaying very broad and diffuse bubble-like lobes. It is 
relatively nearby with a redshift of 0.0857. Miley \etal\ (1981) first 
detected optical continuum and emission lines associated with the extended
radio lobes. There is an extensive network of emission line filaments which
seem to wrap around the radio lobes \cite{van85,tad00}. Van Breugel \etal\ 
(1985)
suggested that the southern jet is deflected by a massive cloud 
at the location of knot K1. 
Morganti \etal\ (2002) detect the 21 cm line of HI in absorption against
the extended radio lobes of Coma A. They suggest that the source is expanding
into a gaseous disk, presumably acquired in a merger which triggered the
radio activity.  The kinematics of the emission line gas are consistent with 
interaction with the radio source \citep{van85,solo03}.

\noindent
The HST and radio images are given in Figures 5-8. Values of the [OIII]/[OII]
 line ratio are given in Table 6 for selected regions. The medium band 
continuum image shows a very prominent nucleus and the faint continuum 
emission associated with the northern knot K1(Fig. \ref{277_cont_radio}). 
The peak of the 
continuum emission is shifted just to the east of the radio peak of K1. 
In our narrow band LRF images we detect 
the emission line gas associated with the nucleus and the filament which runs
 mainly N-S and to the East of the two radio knots 
(Figs. \ref{277_OIII_radio}, 
\ref{277_OII_radio}).  We do not detect any line emission directly associated
 with knot K2. We see a peak in the line emission just to the east of knot
 K1. The line emission in this region exceeds that from the nucleus. The observed
 [OIII]/[OII] ratio also peaks just east of K1 with a relatively high
value of $\sim 5.4$ consistent with the suggestion of Solorzano-I\~{n}arrea 
\& Tadhunter (2003)
that the emission lines in this feature are dominated by photoionized
precursor emission. 

\noindent
 The linear structures visible in the high resolution radio images are very
interesting. They are suggestive of shocks that have been
driven back into the jets as a consequence of the strong interaction with
the emission line cloud. The fact that they are not exactly perpendicular
to the jets, but misaligned by $\sim$30 degrees to perpendicular, may be
significant, since oblique shocks have been proposed as a mechanism for
deflecting jets, and the jet is indeed significantly deflected in
3C277.3. The modeling work of Tingay (1997) demonstrates that it is
possible to produce a significant deflection  for an obliquity of ~30
degrees, although as he notes, it may be difficult to produce a deflection
as large as that seen in 3C277.3, unless the jet is inclined towards the
observer, so that the real jet deflection is less than measured.

\noindent
The nature of the jet-cloud interaction appears somewhat different to
those associated with the other two sources. Whereas in 3C171 and
PKS2250-41 the {\it lowest} ionization features are closely associated with
the detailed radio features (knots, hot-spots), in 3C277.3 the highest
ionization feature {\it within the cloud} is closely associated with the radio 
knot. While this
close association -- within 0.2 arcseconds in our HST observations --
suggests that the jet is ionizing the cloud, it is significant that,
unlike the other two sources and also the extended radio
lobe in 3C277.3 itself \cite{solo03}, the line emission
associated with the jet-knots in 3C277.3 shows no sign of kinematic
disturbance. All this is consistent with the idea that, in this case, we
are observing pure shock photoionized precursor emission; the shock
driven through the warm
ISM may be so fast that the shock gas has not yet cooled sufficiently to
radiate optical emission lines, or the post shock cloudlets may have been
shredded/dispersed by hydrodynamic interaction with the hot post-shock wind
before they had a chance to cool (e.g., Klein, McKee \& Colella 1994).
This may be testable with X-ray observations.
However, we cannot entirely rule out the idea the precursor gas in the
cloud is photoionized by a narrow blazar beam from the AGN which points 
exactly in the jet direction; the ionization gradients across the
cloud are consistent with this idea provided that the Lorentz factor of the
jets close to the nucleus is fairly high (in order to give a narrow beam).

\noindent
The optical continuum emission associated with the warm emission line
cloud is also interesting. Miley \etal\ (1981) detected significant
optical polarization ($14 \pm 3 \%$ with a position angle of 
$71\deg \pm 6\deg$) 
that is aligned perpendicular to the radio axis.
On the basis of the optical/radio spectral index and
polarization they interpreted the optical continuum as
synchrotron emission from the same knots that emit the radio emission.
However, in this case it is surprising that the continuum emission
visible in our HST images appears to be significantly displaced to the
east of peak of the high frequency radio emission. On the 
other hand, the polarization properties of the
continuum emission, and its morphological similarity with the [OIII]
emission, are consistent with the idea that it represents light from the
blazar beam of the AGN scattered into our line of sight (very similar to
PKS2152-69: see Tadhunter \etal\ 1988, di Serego Alighieri \etal\ 1988).
This would also support the idea that the cloud is photoionized by the
blazar beam.

\subsection{PKS 2250-41}

\noindent
PKS 2250-41 is a FRII source at a redshift of 0.308. The first detailed 
observations by \cite{tad94} showed large scale emission line 
arcs or filaments associated with the radio lobes. Both the kinematics 
and the ionization of the gas appear consistent with a scenario 
in which the radio source is interacting with the emission line gas 
\citep{cla97,vill99}. Clark \etal\ (1997) find that the radio emission 
from the western lobe has very low 
polarization consistent with depolarization by surrounding/mixed gas. 

\noindent
Our 15 GHz VLA observations resolve structure in the lobes and reveal 
evidence for secondary hot-spots (e.g., Fig. \ref{pks_OII_radio}). We also
(just) detect the nucleus in the radio which allows improved registration 
of the optical and radio images. 

\noindent
The HST and radio images are given in Figures 9-12. Values of the 
[OIII]/[OII] line ratio are given in Table 6 for selected regions. Note that 
unfortunately  the location of the redshifted wavelength of the 
[OII] line on the LRF lies close to the edge of the chip. Thus, 
the coverage of the [OII] line does not include the Eastern lobe. 
This may not be significant as ground based imaging by Clark \etal\ (1997) 
does not show any emission line gas associated with this lobe. 

\noindent
Our LRF images show line 
emission from the nuclear region and from the bright filament which 
points toward the companion galaxy to the north-east. There is also a hint of 
much more diffuse line emission which is not well detected in these 
observations. As noted by Tadhunter \etal\ (1994), the emission line gas in the Western 
lobe takes the form of an arc which wraps around the outer boundary 
of the lobe (Fig. \ref{pks_OIII_radio}). The [OIII] extends 
in a roughly  180$^\circ$ semi circle around the front side of the 
lobe. The [OII] is somewhat smaller in extent and is concentrated closer 
to the source axis and appears ``nested" within the [OIII] arc. The 
[OII] emission from this 'nested' region exceeds that from the nucleus. 
The [OIII]/[OII] ratio in the lobe is low ($\lae 0.5$) and is somewhat higher
 at the leading edge of the lobe and declines slightly with distance inwards 
towards the nucleus. 

\noindent
The differences between the detailed [OIII] and [OII] emission
line morphologies in the lobe can be explained in terms of different 
proportional contributions of precursor, AGN-photoionized, and 
lower ionization, kinematically-disturbed post-shock gas, with 
AGN photoionization dominating close to the nucleus, 
shock (photo)ionization of the precursor emission dominating in the 
extended arc, and cooling post-shock gas dominating the [OII] emission in 
the lobe. This is consistent with the interpretation of Clark \etal\ 
(1997) and Villar-Mart\'{\i}n \etal\ (1999).

\noindent
The medium band continuum image shows a possible companion galaxy to the 
north (Clark \etal\, 1997) and emission associated with the secondary hot-spot 
in the western lobe. A spur of the continuum emission appears to point 
toward the companion while the companion itself is elongated in the 
direction of the nucleus. The position angles of the spur in the continuum, 
appears to be aligned with the major axis of the companion to within 6 
degrees. The alignment is even better between the PA for the major axis of the 
companion and the filament of PKS2250-41. Table 5 summarizes the position 
angles of each of the distinctive features of the companion and PKS2250-41. 
Recent VLT observations of the putative companion galaxy show Balmer 
absorption lines at the same redshift as PKS2250-41 indicating that it is 
indeed a companion in the same group and that it has had recent star 
formation (Tadhunter, in preparation). 

\noindent
The spatial coincidence between the optical continuum emission and secondary
hot-spot in the western lobe is striking. One possibility is that the secondary
hot-spot in fact represents compact radio emission from the nucleus of a
 companion galaxy that the jet is colliding with (consistent with the
jet-galaxy collision idea presented in Clark \etal\ 1997). In this case,
the optical light represents the stellar light of the bulge of this companion
galaxy to the west. Alternatively, both the optical and radio emission might
be explained in terms of synchrotron emission from the secondary hot-spot.
However this possibility is ruled out as a dominant source of emission,
due to the lack of optical polarization ($\lae 4.5\%$, see Shaw \etal\ 1995).

\noindent
The highly collimated nature of the emission line structure in the nuclear 
region is reminiscent of the structures aligned along the radio axis in high
redshift radio galaxies (Best \etal\, 2000), yet in this case the structure 
is misaligned from the radio axis by ~25 degrees. While radio and optical 
structures in high z radio galaxies are rarely in perfect alignment, 
 a deviation of this size is seen in less than 25\% of 
powerful radio sources studied by Chambers, Miley \& van Breugel (1987). 
Moreover, in contrast to the high-z 
objects, the emission line kinematics of the linear 
structure are quiescent. However, the fact that this structure  points 
towards the north-eastern companion (and the companion points towards it!), 
suggests that the gas has been captured in an interaction with the companion, 
and that the linear structure is a result of this interaction (perhaps a disk 
of captured gas seen edge on). In contrast, on the western side of the nucleus 
in PKS2250-41 the jets are likely to be directly interacting with the 
debris of the other companion galaxy. Therefore one can interpret PKS2250-41 
as a complex system in which the dominant galaxy is interacting with two 
companions, and the jet is ploughing into the debris of the companion on the 
west side of the nucleus.

\section{CONCLUSIONS}

\noindent
The HST imaging results presented in this paper demonstrate the full
diversity of jet-cloud interaction phenomenon, with structures ranging
from the bow-shock-like structure that cocoons the western radio lobe in
PKS2250-41, to the linear structures closely aligned along the radio axis
in 3C171, to the bright emission line cloud associated with the knot along
the radio jet in 3C277.3. Amongst low/intermediate redshift radio galaxies 
PKS2250-41, 3C277.3 and 3C171 are unusual in the strengths of their 
jet-cloud interactions. In addition, PKS2250-41 and 3C171 have close 
companions aligned along their radio jets while 3C277.3 shows evidence of 
a recent merger in which the disk of the debris is observed face-on. It is
likely that these objects represent the few rare cases in which the
jets are ploughing into the debris left over from the merger events
that triggered the activity.

\noindent
Significant differences are seen between the [OII] and [OIII] structures 
in all three objects. All of these differences can be explained in terms
of varying contributions of cooling post-shock gas, shock photoionized
precursor gas, and AGN photoionized gas.
The correlations between the emission line morphology and 
the [OIII]/[OII] line ratios with the radio emission seen in ground based
observations are clarified by the HST imaging. We confirm that the radio lobes
and hot-spots are preferentially associated with lower ionization gas. 
3C 171 exhibits high surface brightness emission line gas mainly along 
the radio source axis. The lowest ionization gas is seen at the Eastern hot
spot.  
In 3C 277.3 there is bright high ionization gas (and continuum) offset just 
to the 
east of  the  radio knot K1. Our observations are consistent with the 
suggestion of \cite{solo03} that this emission is produced by precursor gas
ionized by the shock being driven into the cloud by the deflected radio jet. 
In PKS 2250-41 we resolve the emission line arc which wraps around the outer 
rim of the western lobe. The lower ionization [OII] emission is nested
just interior to the higher ionization [OIII] emission, suggesting that
we have resolved the cooling region behind the bow shock. We also detect
possible continuum emission from the secondary hot-spot. 

\noindent
Thus, our observations support the hypothesis that in these sources, the
interaction between the expanding radio source and the ambient gas strongly
influences the morphology, kinematics,  and ionization of the gas. 

\section{Acknowledgements}
\acknowledgments
\noindent
Support for program 6657 (PI C. Tadhunter) was provided by NASA through a 
grant from the Space Telescope Science Institute, which is operated by the 
Association of Universities for Research in Astronomy, Inc., under NASA 
contract NAS 5-26555.  The National Radio Astronomy Observatory is a facility 
of the National Science Foundation operated under cooperative agreement 
by Associated Universities, Inc. MERLIN is operated as a National Facility 
by the University of Manchester on behalf of the UK Particle Physics and 
Astronomy Research Council. This research made use of (1) the NASA/IPAC 
Extragalactic Database(NED) which is operated by the Jet Propulsion Laboratory,
 California Institute of Technology, under contract with the National 
Aeronautics and Space Administration; and (2) NASA's Astrophysics Data 
System Abstract Service. We thank Shireen Gonzaga for help with the 
WFPC2 observations and Peter Thomasson and Tom Muxlow for 
assistance with the MERLIN2 observations for 3C277.3. Karen Wills 
acknowledges financial support from PPARC and the Royal Society. We thank the
anonymous referee for very detailed and helpful comments.

%\newpage
\clearpage

\clearpage

\clearpage

\begin{figure}
\epsscale{0.85}
%\plotone{3C171ratio_radio1.eps}
\plotone{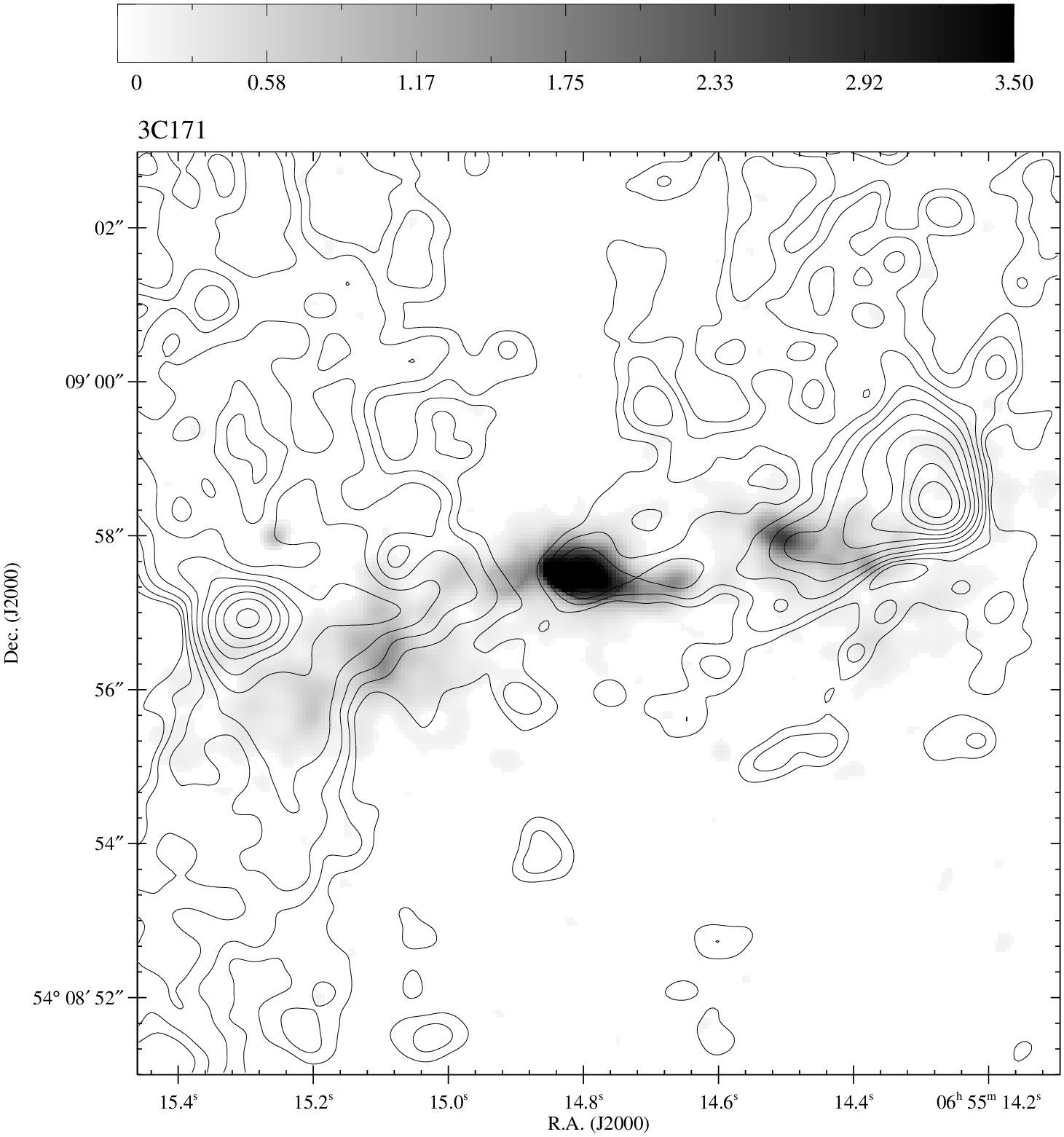}
\caption{3C171: Contour plot of radio image overlaid on gray-scale plot of
 the [OIII]$\lambda$5007/[OII]$\lambda$3727 ratio image. 
 Non-detections of [OII] are 
replaced with the 3$\sigma$ value to allow displays of lower limits to the 
ratio. 
The 8 GHz VLA radio image is displayed with contour levels
starting at 3 times the rms noise:
3$\times$2.167$\times$$10^{-5}$$\times$[1, 2, 4, 8, 16, 32, 64, 128, 
256, 512, 1024, 2048]Jy/beam. \label{171_ratio_radio1}
}
\end{figure}

\clearpage

\begin{figure}
\epsscale{0.85}
%\plotone{3C171cont_radio.eps}
\plotone{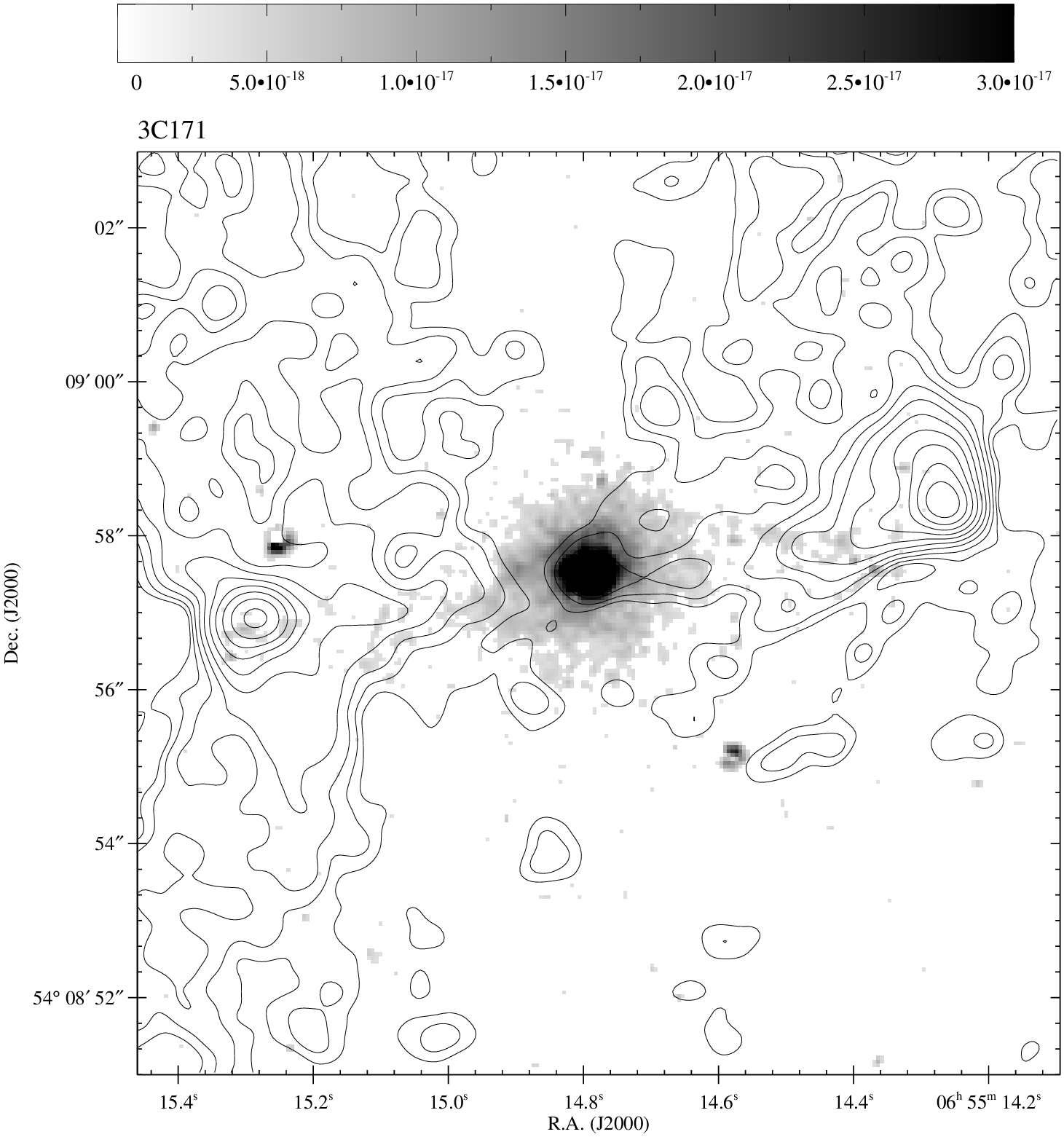}
\caption{3C171: Contour plot of radio image overlaid on gray-scale F547M 
continuum image. The continuum image has been masked to retain emission above 
3.651$\times$$10^{-18}$ ergs  s\mone\ cm\mtwo\ arcsec\mtwo.
The 8 GHz VLA radio image is displayed with contour levels
starting at 3 times the rms noise:
3$\times$2.167$\times$$10^{-5}$$\times$[1, 2, 4, 
8, 16, 32, 64, 128, 256, 512, 1024, 2048]Jy/beam. \label{171_cont_radio}
}
\end{figure}

\clearpage

\begin{figure}
\epsscale{0.85}
%\plotone{3C2773_overlay.eps}
\plotone{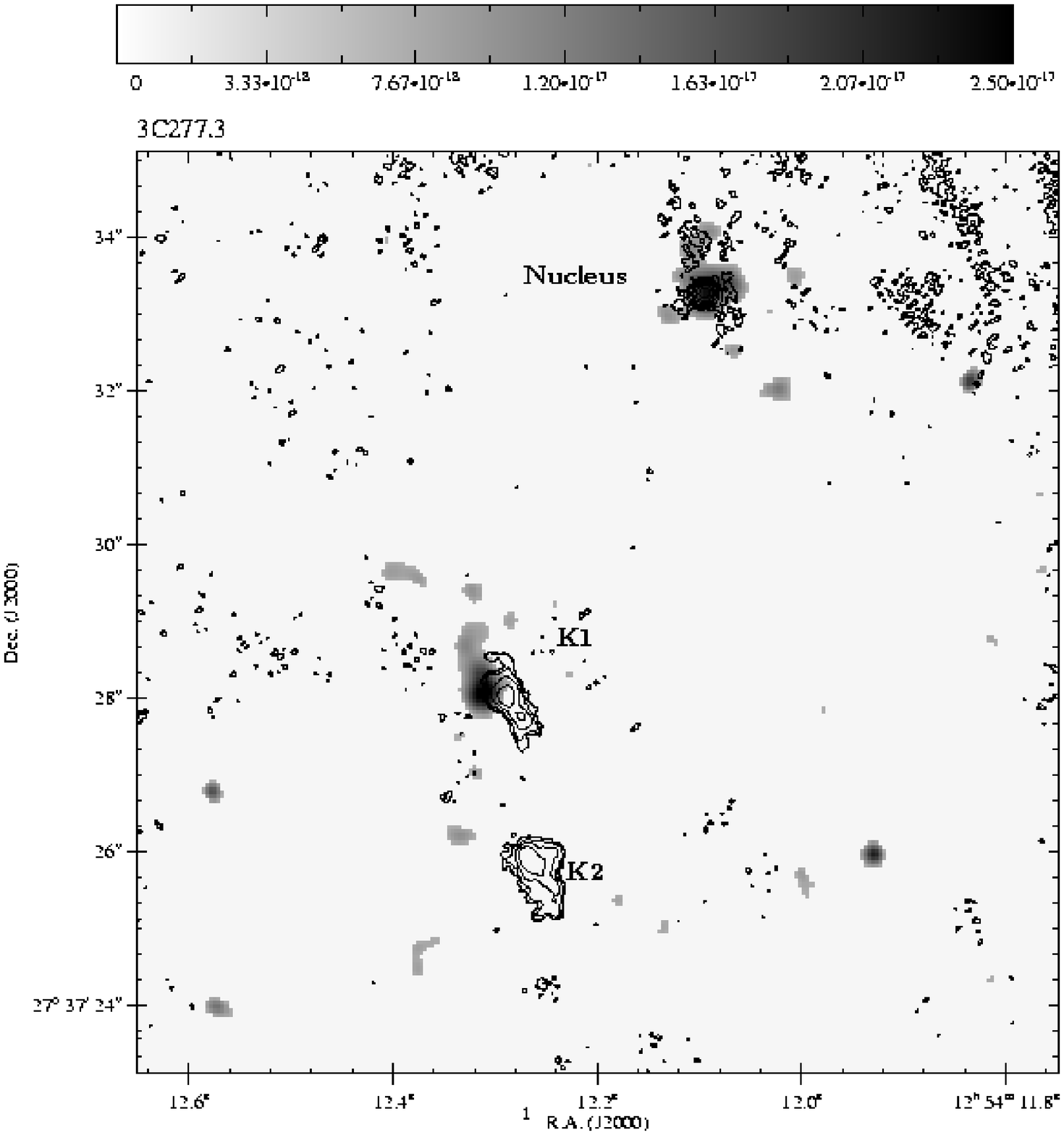}
\caption{3C277.3: Contour plots of radio image are overlaid on gray-scale surface 
brightness plot of [OII]$\lambda$3727 emission line image. The [OII] image
has been masked to retain emission above 7.332$\times$$10^{-18}$ ergs s\mone\
cm\mtwo\ arcsec\mtwo.
The 4.8 GHz MERLIN2 + VLA radio image is  displayed with contour levels
starting at 3 times the rms noise:
3$\times$1.608$\times$$10^{-5}$
$\times$[1, 2, 4, 8, 16, 32, 64, 128, 256, 512, 1024, 2048]Jy/beam. \label{277_OII_radio}
}
\end{figure}

\clearpage

\begin{figure}
\epsscale{0.85}
%\plotone{3C2773OIII_radio.eps}
\plotone{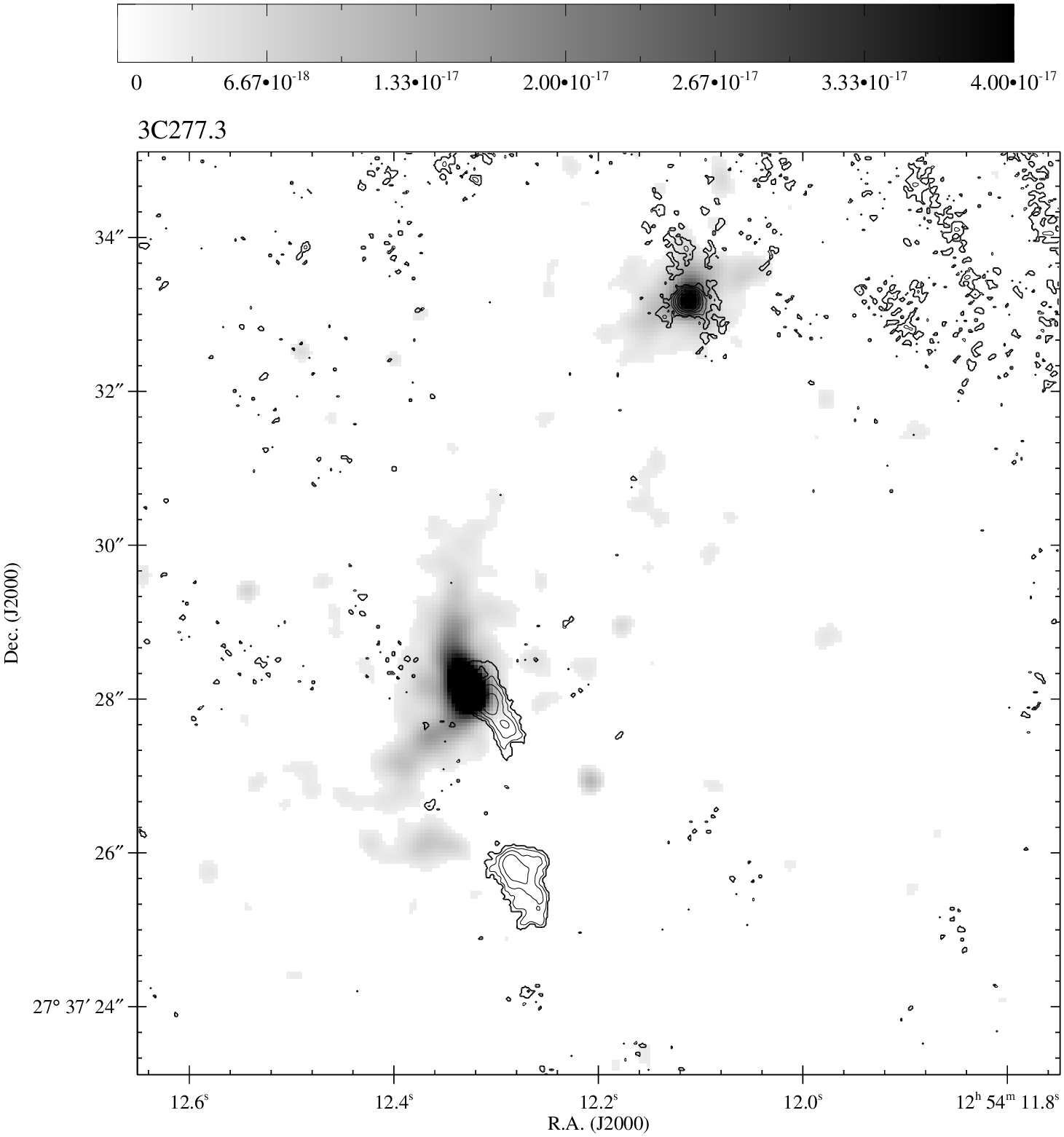}
\caption{3C277.3: Contour plots of radio image are overlaid on gray-scale surface 
brightness plot of [OIII]$\lambda$5007 emission line image. The [OIII] image
has been masked to retain emission above 1.033$\times$$10^{-18}$ ergs s\mone\
cm\mtwo\ arcsec\mtwo.
The 4.8 GHz MERLIN2 + VLA radio image is  displayed with contour levels
starting at 3 times the rms noise:
3$\times$1.608$\times$$10^{-5}$
$\times$[1, 2, 4, 8, 16, 32, 64, 128, 256, 512, 1024, 2048]Jy/beam. 
\label{277_OIII_radio}
}
\end{figure}

\clearpage
\begin{figure}
\epsscale{0.85}
%\plotone{3C2773ratio_radio1.eps}
\plotone{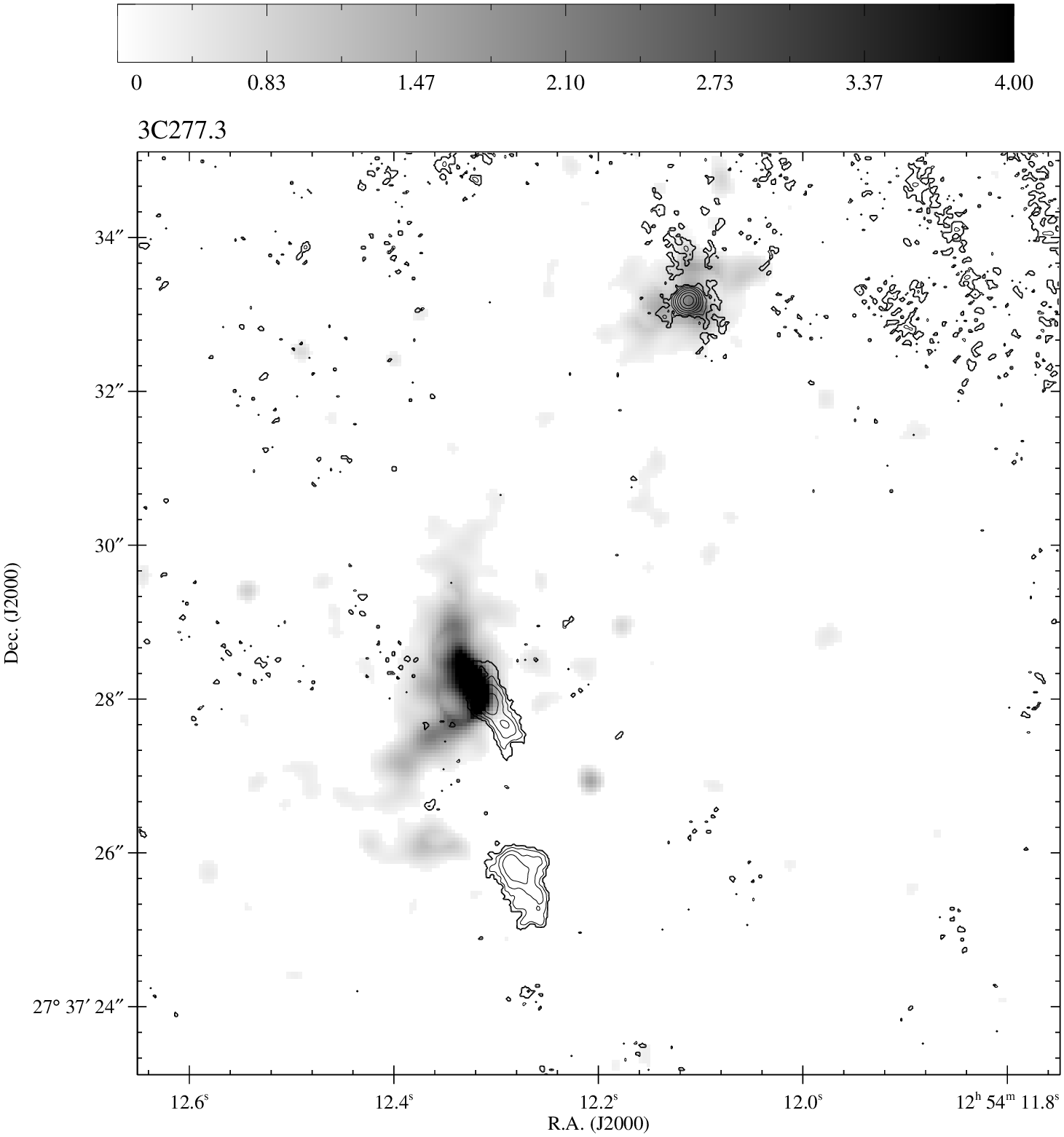}
\caption{3C277.3: Contour plot of radio image overlaid on gray-scale plot of 
the [OIII]$\lambda$5007/[OII]$\lambda$3727 ratio image. Non-detections of [OII] are 
replaced with the 3$\sigma$ value to allow displays of lower limits to the 
ratio. 
The 4.8 GHz MERLIN2 + VLA radio image is  displayed with contour levels
starting at 3 times the rms noise:
3$\times$1.608$\times$$10^{-5}$$\times$[1, 2, 4, 8, 16, 32, 64, 128,
 256, 512, 1024, 2048]Jy/beam. \label{277_ratio_radio1}
}
\end{figure}

\begin{figure}
\epsscale{0.85}
%\plotone{3C2773cont_radio.eps}
\plotone{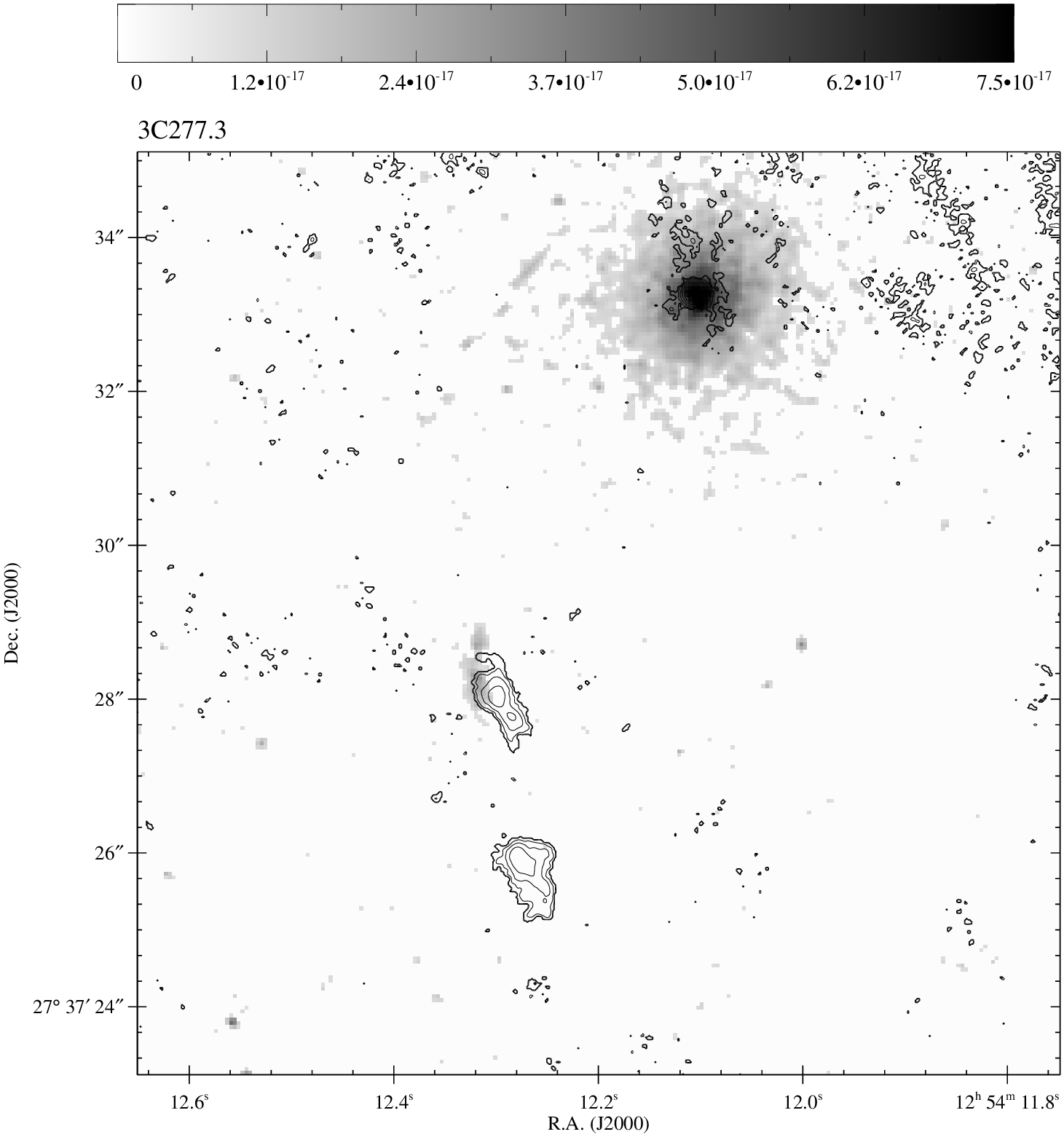}
\caption{3C277.3: Contour plots of radio image are overlaid on gray-scale 
F467M continuum
 image. The continuum image has been masked to retain emission above 
8.931$\times$$10^{-18}$ ergs s\mone\ cm\mtwo\ arcsec\mtwo.
The 4.8 GHz MERLIN2 + VLA radio image is  displayed with contour levels
starting at 3 times the rms noise:
3$\times$1.608$\times$$10^{-5}$$\times$[1, 2, 4, 8, 
16, 32, 64, 128, 256, 512, 1024, 2048]Jy/beam. \label{277_cont_radio}
}
\end{figure}

\begin{figure}
\epsscale{0.85}
%\plotone{PKS2250-41OII_radio.eps}
\plotone{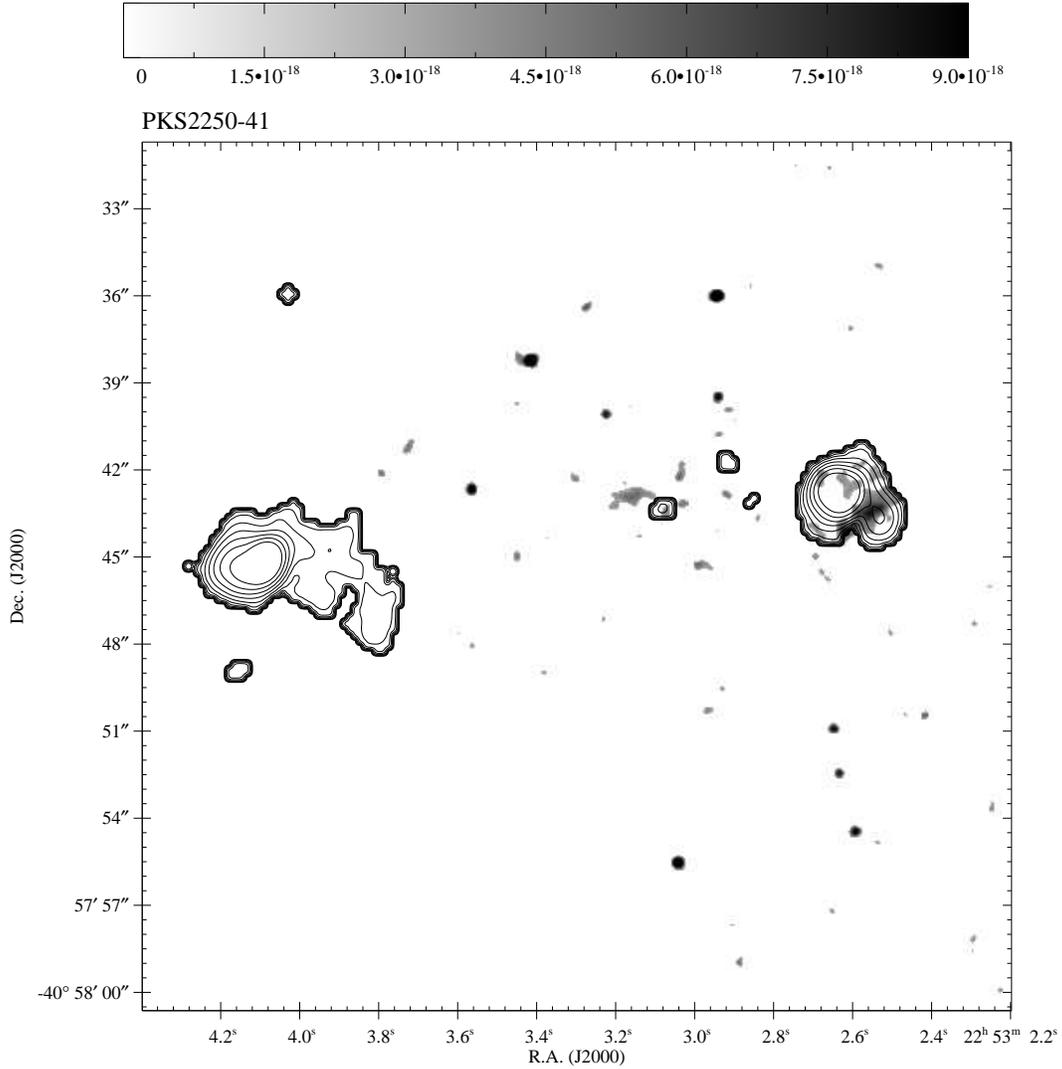}
\caption{PKS2250-41: Contour plot of radio image overlaid on gray-scale surface 
brightness plot of [OII]$\lambda$3727 emission line image. The [OII] image
has been masked to retain emission above 3.156$\times$$10^{-18}$ ergs s\mone\
cm\mtwo\ arcsec\mtwo. 
The eastern lobe detected in the radio image is off the edge of the [OII] image. 
The 15 GHz VLA radio image is  displayed with contour levels
starting at 3 times the rms noise:
3$\times$3.180$\times$$10^{-6}$
$\times$[1, 2, 4, 8, 16, 32, 64, 128, 256, 512, 1024, 2048]Jy/beam. 
\label{pks_OII_radio}
}
\end{figure}

\clearpage

\begin{figure}
\epsscale{0.85}
\plotone{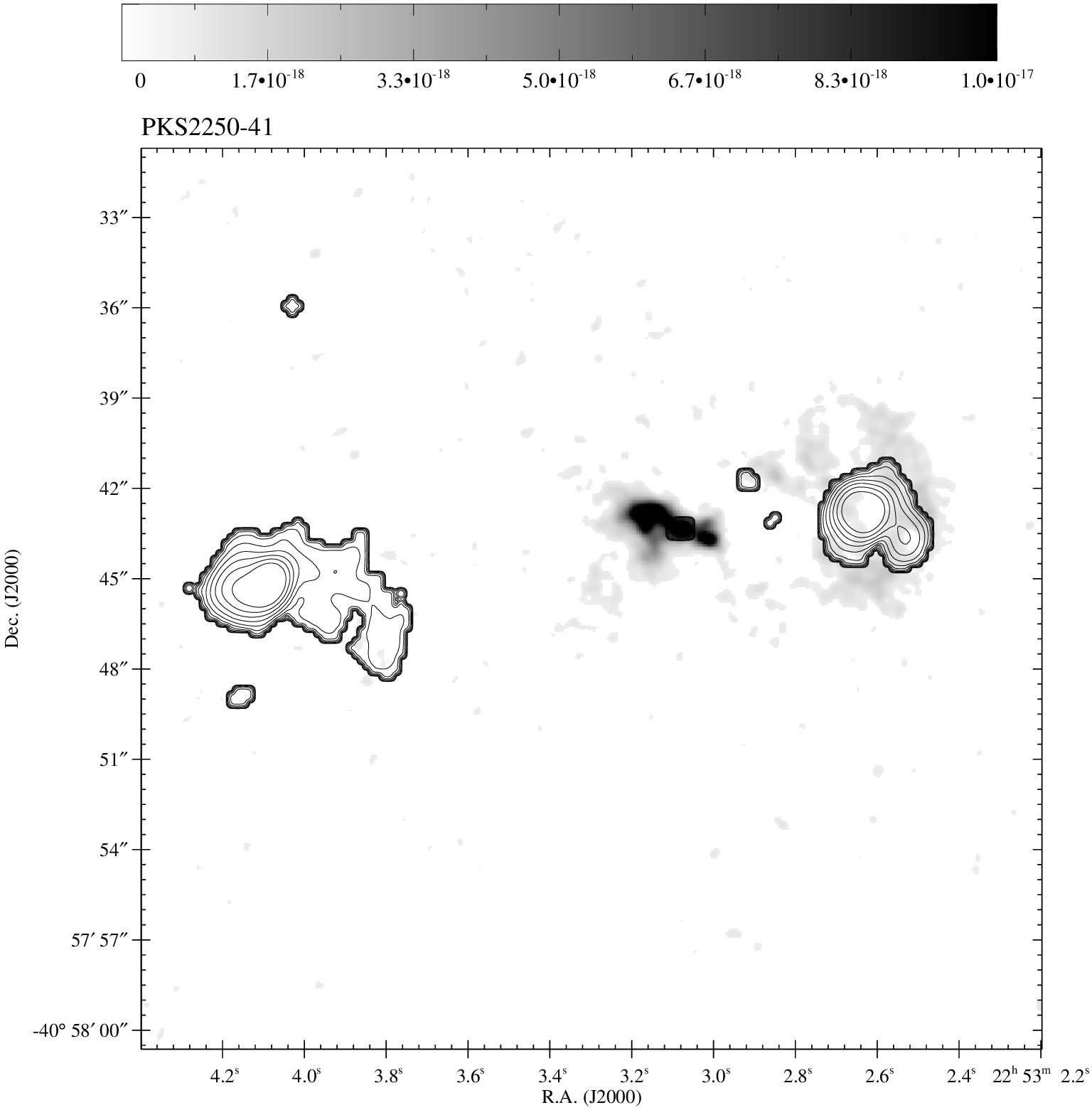}
%\plotone{PKS2250-41OIII_radio.eps}
\caption{PKS2250-41: Contour plot of radio image overlaid on gray-scale surface 
brightness plot of [OIII]$\lambda$5007 emission line image. The [OIII] image
has been masked to retain emission above 6.657$\times$$10^{-19}$ ergs s\mone\
cm\mtwo\ arcsec\mtwo.
The 15 GHz VLA radio image is  displayed with contour levels
starting at 3 times the rms noise:
3$\times$3.180$\times$$10^{-6}$
$\times$[1, 2, 4, 8, 16, 32, 64, 128, 256, 512, 1024, 2048]Jy/beam. 
\label{pks_OIII_radio}
}
\end{figure}

\begin{figure}
\epsscale{0.85}
%\plotone{PKS2250-41ratio_radio1.eps}
\plotone{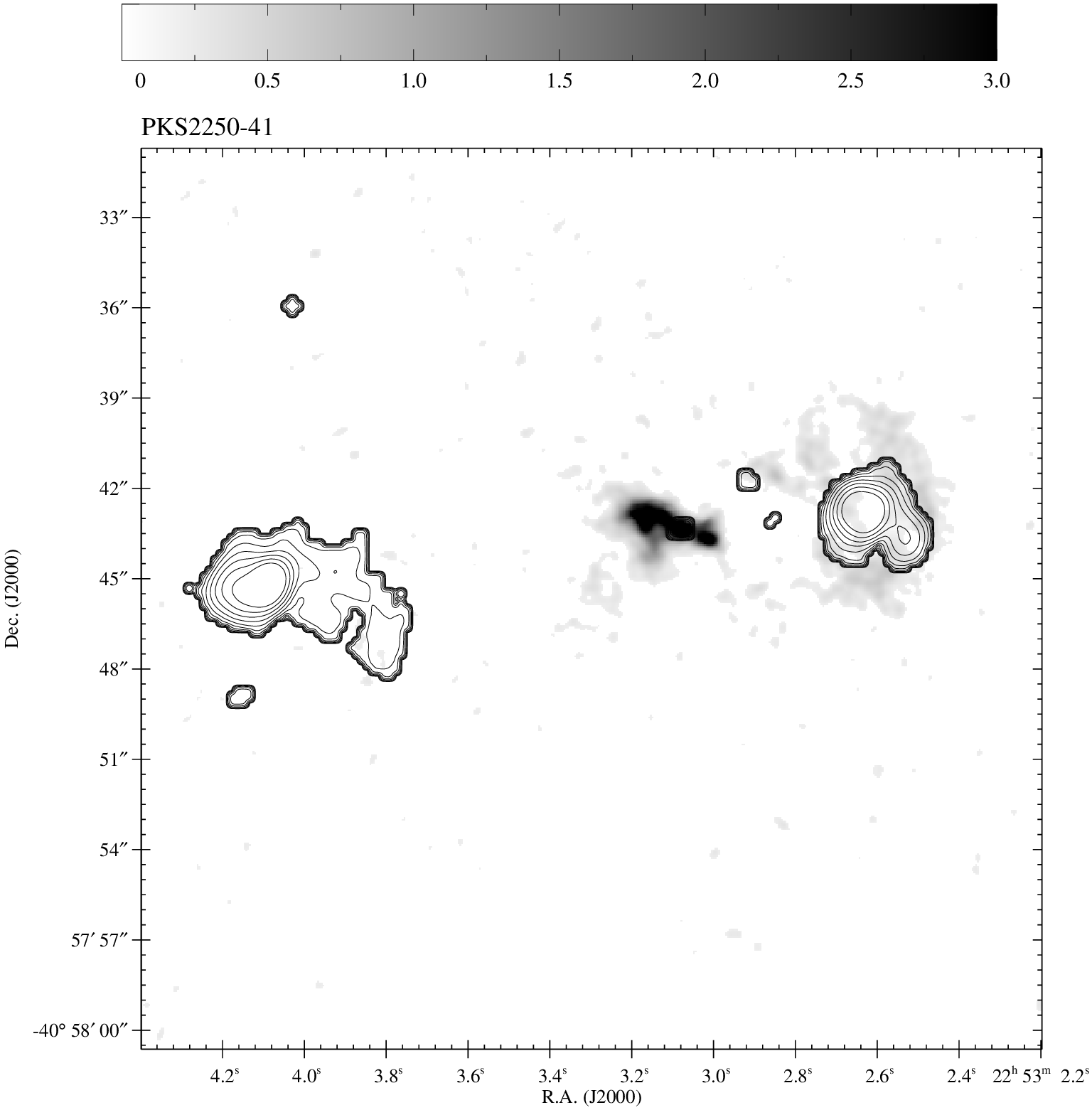}
\caption{PKS2250-41: Contour plot of radio image was overlaid on gray-scale surface
 plot of the [OIII]$\lambda$5007/[OII]$\lambda$3727 ratio image. 
Non-detections of 
[OII] are replaced with the 3$\sigma$ value to allow displays of lower limits 
to the ratio. 
The 15 GHz VLA radio image is  displayed with contour levels
starting at 3 times the rms noise:
3$\times$
3.180$\times$$10^{-6}$$\times$[1, 2, 4, 8, 16, 32, 64, 128, 256, 512,
 1024, 2048]Jy/beam. \label{pks_ratio_radio1}
}
\end{figure}

\begin{figure}
\epsscale{0.85}
%\plotone{PKS2250-41_overlay.eps}
\plotone{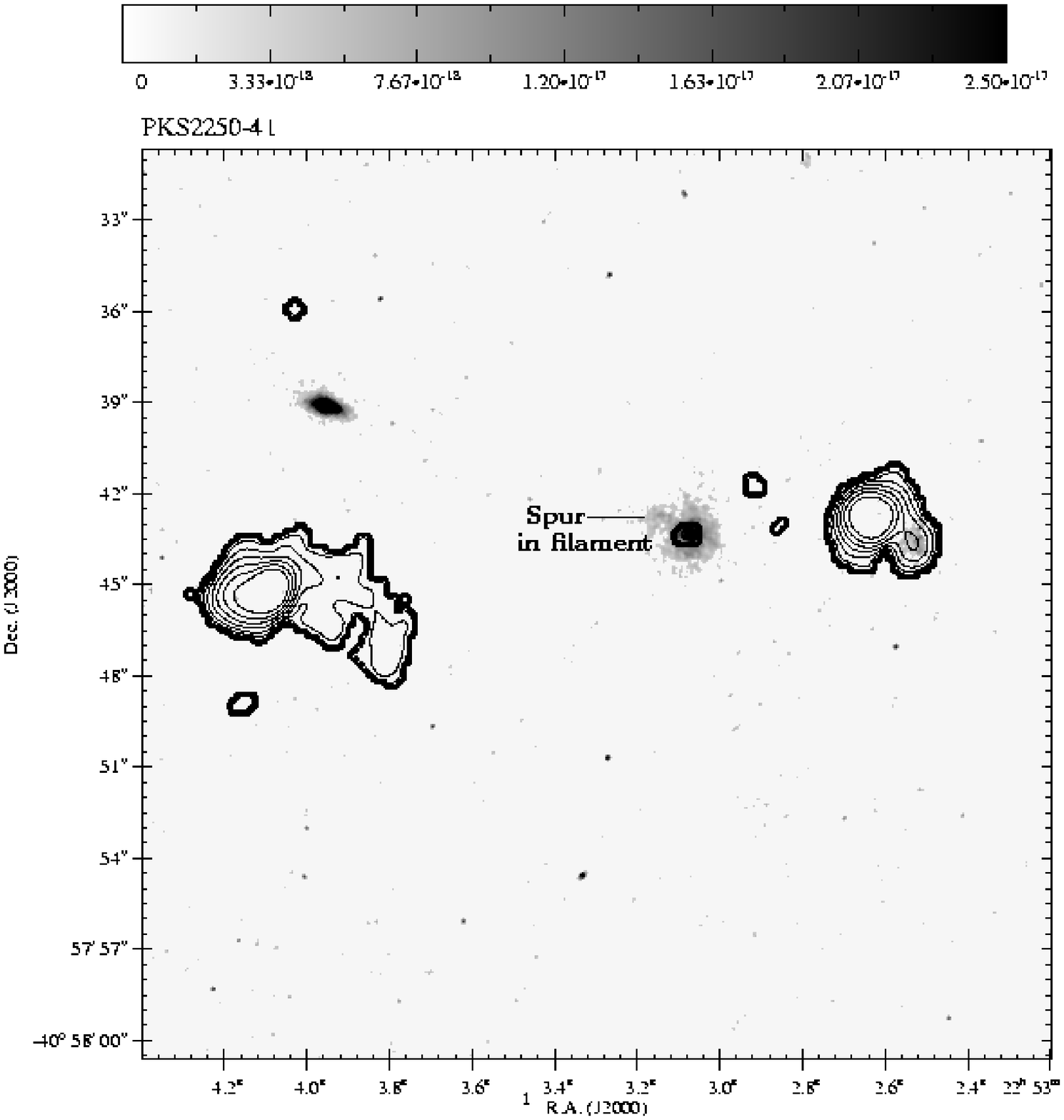}
\caption{PKS2250-41: Contour plot of radio image overlaid on gray-scale F547M 
continuum image. The continuum image has been masked to retain emission above 
3.858$\times$$10^{-18}$ ergs s\mone\ cm\mtwo\ arcsec\mtwo.
The 15 GHz VLA radio image is  displayed with contour levels
starting at 3 times the rms noise:
3$\times$3.180$\times$$10^{-6}$$\times$[1, 
2, 4, 8, 16, 32, 64, 128, 256, 512, 1024, 2048]Jy/beam. 
\label{pks_cont_radio}
}
\end{figure}

\clearpage

\begin{deluxetable}{crrrrrr}
\tabletypesize{\scriptsize}
\tablecaption{Properties of the galaxies \label{Table 1}}
\tablewidth{0pt}
\tablehead{
\colhead{Object} &\colhead{RA(J2000)}   &\colhead{DEC(J2000)}   &
\colhead{redshift\tablenotemark{a}} & \colhead{Magnitude} &\colhead{E(B-V)} &\colhead{Applied correction\tablenotemark{c}}\\ &h m s &d m s & &mag &mag &\\(1) &(2) &(3) &(4) &(5) &(6) &(7)\\ }
\startdata
3C171 &06 51 14.8 &$+$54 09 00 &0.2384 &18.89\tablenotemark{1} &0.40\tablenotemark{b} &0.87\\
3C277.3 &12 54 11.7 &$+$27 37 33 &0.0857 &15.94\tablenotemark{1} &0.012\tablenotemark{a} &0.99\\
PKS2250-41 &22 53 03.1 &$-$40 57 46 &0.3100 &20.0\tablenotemark{2} &0.14\tablenotemark{b} &0.95\\
\enddata
\tablenotetext{1}{V band magnitude, from Spinrad \etal\ (1985)}
\tablenotetext{2}{m$_{\rm p}$ from  Hunstead (1971)}
\tablenotetext{a}{Obtained from NASA Extragalactic Database}
\tablenotetext{b}{Determined using observed H$\alpha$/H$\beta$ ratio \citep{cla97,cla98}}
\tablenotetext{c}{Total reddening correction using E(B-V) in Column (6) applied to 
the [OIII]/[OII] ratio image.}
\end{deluxetable}

%\clearpage

\begin{deluxetable}{crrrrrr}
\tabletypesize{\scriptsize}
\tablecaption{Radio Properties \label{Table 2}}
\tablewidth{0pt}
\tablehead{
\colhead{Object} &\colhead{Radio Power}\tablenotemark{a}   &\colhead{Radio Luminosity}   &\colhead{Angular Size\tablenotemark{b}} &\colhead{Linear Size} &\colhead{Scale} &\colhead{Spectral Index}\\
 &ergs/s/Hz &ergs/s &arcsec &kpc &kpc/arcsec}

\startdata
3C171 &5.18$\times$10$^{33}$ &1.39$\times$10$^{43}$ &33\tablenotemark{c} &111.5 &3.38 &-0.94\\
3C277.3 &4.56$\times$10$^{32}$ &1.14$\times$10$^{41}$ &40 &58.8 &1.47 &-0.70\\
PKS2250-41 &2.75$\times$10$^{33}$ &2.06$\times$10$^{43}$ &77  &21.4 &4.05 &-0.99\\

\enddata
\tablecomments{The references for spectral index are 3C171: Clark \etal\ (1998);
3C277.3: van Breugel \etal\ (1985); PKS2250-41: Morganti \etal\ (1993).}
\tablenotetext{a}{Corresponding frequency for 3C171 and 3C277.2 is 1.4GHz and for PKS2250-41 is 4.885GHz}
\tablenotetext{b}{Refers to the maximum extension of the radio source}
\tablenotetext{c}{Includes the diffuse plumes}
\end{deluxetable}.

\clearpage

\begin{deluxetable}{crrrrrrr}
\tabletypesize{\scriptsize}
\tablecaption{HST Optical images \label{Table 3}}
\tablewidth{0pt}
\tablehead{
\colhead{Object} &\colhead{Emission Line}   &\colhead{Chip}   &
\colhead{Filter} & \colhead{Date} &\colhead{Exposure Time} &\colhead{Mean $\lambda$} &\colhead{Bandwidth}\\ & & & & & sec &\AA &\AA}
\startdata
 &[OII] &3 &FR418N &02/10/1999 &2000 &4615 &60 \\
3C171 &[OIII] &2 &FR680P15 &02/10/1999 &3000 &6195 &81\\
 &Continuum &2 &F547M &02/10/1999 &1000 &5484 &484\\\hline
 &[OII] &3 &FR418N18 &08/01/1997 &2000 &4044 &53\\
3C277.3 &[OIII] &4 &FR533N &08/01/1997 &1000 &5436 &71\\
 &Continuum &2 &F467M &08/01/1997 &1000 &4670 &178\\\hline
 & [OII] &3 &FR533N &09/01/1999 &2000 &4862 &63\\
PKS2250-41 &[OIII] &3 &FR680N &09/01/1999 &3000 &6550 &85\\
 &Continuum &2 &F547M &09/01/1999 &1000 &5484 &484\\

\enddata
\tablecomments{The pixel scale for Chip 2 is 0.9961 arcsec/pixel, for chip 3 is 0.9958 
arcsec/pixel and for chip 4 is 0.9964 arcsec/pixel. The observations are from progam 6657.
The bandwidth for the LRF images is $\sim 1.3\%$ of the central wavelength. }

\end{deluxetable}

\clearpage

\begin{deluxetable}{crrrrrrr}
\tabletypesize{\scriptsize}
\tablecaption{Properties : Radio images \label{Table 4}}
\tablewidth{0pt}
\tablehead{
\colhead{Object} &\colhead{Telescope}   &\colhead{Date}   &
\colhead{Integration time} & \colhead{Frequency} &\colhead{Bandwidth} 
&\colhead{rms} &\colhead{Clean Beam }
\\ & & & & & &mJy/beam & }
\startdata
3C171 &VLA &08/06/1995 &3550s &8 GHz &50 MHz &0.044 &$0.35\asec \times 0.25\asec$ PA= $75\deg$ \\
3C277.3 &VLA\tablenotemark{a} &06/10/1982 &680s & 5 GHz & 50 MHz & \\
3C277.3 &Merlin2 &03/04/2000 &36 hrs &5 GHz &15 MHz &0.043\tablenotemark{b} &$0.15\asec \times 0.15\asec$ \\ 
PKS2250-41 &VLA &09/17/1995 &5210s &15 GHz &50 MHz &0.090 &$0.82\asec \times 0.33\asec$ PA= $-11.4\deg$ \\
\enddata
\tablenotetext{a}{VLA archive data from J. Ekers(1982).}
\tablenotetext{b}{Combined MERLIN2 and VLA data.}

\end{deluxetable}

\clearpage

\begin{deluxetable}{crrrr}
\tabletypesize{\scriptsize}
\tablecaption{Properties : Position angles in PKS~2250-41 \label{Table 5}}
\tablewidth{0pt}
\tablehead{
\colhead{Object} &\colhead{Feature} &\colhead{Emission} &\colhead{Position Angle}
\\ & & & (degrees)
}
\startdata
Companion &Orientation vector\tablenotemark{a} &Continuum &71\\
Companion &Major axis &Continuum &69\\
PKS2250-41 & 1\arcsec Spur in Filament &Continuum &63\\
PKS2250-41 & Filament &[OIII] &67\\
\enddata

\tablenotetext{a}{A line joining nuclei of the companion and PKS2250-41}

\end{deluxetable}

\clearpage

\begin{deluxetable}{crrrrrrrr}
\tabletypesize{\scriptsize}
\tablecaption{[OIII]/[OII] Values in Selected Regions \label{Table 6}}
\tablewidth{0pt}
\tablehead{
\colhead{Object} &\colhead{Region} &\colhead{Observed Ratio} \\
\colhead{(1) } &\colhead{(2) } &\colhead{(3)} }
\startdata
3C171 &Nucleus &3.9$\pm$0.6\tablenotemark{a}\\
&Eastern lobe & 1.6$\pm$0.2\\
&Eastern hotspot &$>0.4$\tablenotemark{b}\\
&Western lobe &2.7$\pm$0.4\\
&Western hotspot &$>0.3$\tablenotemark{b}\\\hline
3C277.3 &Nucleus &2.0$\pm$0.3\\ 
&Knot K$_1$ &7.0$\pm$1.0\\
&$1\asec$ E of Knot K$_2$ &1.2$\pm$0.2\tablenotemark{c}\\\hline
PKS2250-41 &Nucleus &3.9$\pm$0.6\\
&Filament &5.7$\pm$0.8\\
&Western hotspot &0.7$\pm$0.1\\
\enddata

\tablecomments{Column 3 gives the maximum  observed values for the 
[OIII]/[OII] ratio in various regions of the galaxies when both 
[OIII] and [OII] are detected with S/N $\gae 3$.} 
\tablenotetext{a}{The uncertainty has been calculated by propagating the 
uncertainties associated with data analysis and photometric calibration
through all the steps in the analysis.}
\tablenotetext{b}{There are regions near the hot spots in 3C171 where the [OII] and
[OIII] emission does not overlap.  
However we do detect [OIII] emission near both hotspots and [OII] at the
Eastern hot spot.  The ratios quoted are for regions near the hotspots 
with significant [OIII] emission but little or no [OII] emission and are 
the lower limit given by setting [OII]  to the 3$\sigma$ value.}
\tablenotetext{c}{Here the quoted value for the ratio corresponds to emission 
$1\asec$ East of K2.}
\end{deluxetable}
\end{document}